\documentclass[
 pra,
 amsmath,amssymb,
 aps,
]{revtex4-2}

\usepackage{graphicx}
\usepackage{dcolumn}
\usepackage{bm}
\usepackage{hyperref}
\usepackage{setspace}

\DeclareMathOperator{\Tr}{Tr}

\begin{document}

\title{
From Laplacian-to-Adjacency Matrix for Continuous Spins on Graphs
}

\author{Nikita Titov}
\email{nikita.titov@units.it}
 \affiliation{Department of Physics, University of Trieste, Strada Costiera 11, 34151 Trieste, Italy}
 
\author{Andrea Trombettoni}
\email{atrombettoni@units.it}
\affiliation{Department of Physics, University of Trieste, Strada Costiera 11, 34151 Trieste, Italy}

\date{\today}

\begin{abstract}
The study of spins and particles on graphs has broad applications, from the dynamics of interacting systems on networks to combinatorial problems.
Here, we study the large-$n$ limit of the $O(n)$ model on graphs, which is considerably more challenging than on regular lattices, as the loss of translational invariance gives rise to an infinite set of saddle point constraints in the thermodynamic limit. We show that the free energy at low and high temperature $T$ is determined by the spectrum of two fundamental graph-theoretic objects: the Laplacian matrix at low $T$ and the Adjacency matrix at high $T$. Their interplay is studied across several classes of graphs. For regular lattices the two coincide. We obtain an exact solution on trees, where the Lagrange multipliers interestingly depend solely on the number of nearest neighbors. We further contrast these classical results with those for a quantum spin model on an exemplary tree.
For decorated lattices, the singular part of the free energy is governed by the Laplacian spectrum, whereas this is true for the full free energy only in the zero temperature limit. Finally, we discuss a bipartite fully connected graph to highlight the importance of a finite coordination number in these results.
\end{abstract}

\maketitle

\section{Introduction }
The assumption of homogeneity is standard practice in theoretical physics, as it enables the use of Fourier transforms and greatly simplifies the analysis of many-body systems \cite{cardy1996scaling,zee2010quantum,simon2013oxford}. This assumption, however, removes the influence of underlying spatial structures. In fact, there are many contexts in which it is necessary to study the properties of particles or spins breaking translational invariance giving rise to new physical phenomena \cite{chaikin1995principles,benavraham2000diffusion,mezard2009information}.
The study of properties of particles or variables on graphs finds applications in diverse areas, spanning from urban and biological networks \cite{barthelemy2022spatial}, to complex systems \cite{albert2002statistical}, random walks \cite{burioni2001random} and statistical mechanics \cite{gefen1980critical,rammal1983random,monceau2011critical,kulske2025ising}. As a further motivation, recent progress in the control of trapping potentials in ultracold atomic systems using digital micromirror devices \cite{gauthier2016direct,smith2021generation} and holographic methods \cite{gaunt2012robust,harte2014conjugate} has enabled the realization of finely tunable lattice geometries and energy potentials, which can also be well controlled by optical lattices \cite{cataliotti2001josephson,morsch2006dynamics} and optical tweezer arrays \cite{kaufman2018microscopic}.
With  holographic techniques a junction of three or four one-dimensional waveguides (forming a so-called $Y$ graph) has been constructed \cite{buccheri2016holographic} and 
similar non-translational invariant topologies have been built by
connecting superconducting Josephson junctions \cite{sodano2006inhomogenous,silvestrini2007topology,lucci2022quantum}. 
These developments allow the direct study of quantum particles and the related effective spin models on controllable non-translational invariant graphs.

Beyond their experimental relevance, the study of models defined on graphs is also connected to combinatorial optimization problems such as Maximum Cut and the Number Partitioning Problem which are generally NP-hard and can be mapped to the ground state energy computation of the Ising model on a complex structure \cite{andrew2014ising}. As an example, the most powerful algorithms for the approximation of Maximum Cut \cite{goemans1995improved} are based on the relaxation of the binary Ising constraints to vectors on a sphere corresponding to $O(n)$ models from statistical mechanics.

The absence of translational symmetry and the interest in studying equilibrium and dynamical properties of particles and spins on graphs motivated the development of numerical and analytical approaches \cite{burioni2000limit,haug2019andreev,tokuno2008dynamics,crampe2013quantum,giuliano2016two}. For instance, belief propagation provides efficient algorithms for classical systems to determine marginals of probability distributions by sending messages from site to site \cite{mezard2009information}. 

A prominent analytical approach is the large-$n$ expansion \cite{moshe2003quantum}, which allows one to approximate both field theories and interacting models on a lattice by a quadratic Hamiltonian subject to a saddle point constraint. It is well known that classical $O(n)$ spin models are equivalent to the so-called spherical model in the large-$n$ limit on regular lattices \cite{berlin1952spherical,lewis1952spherical,stanley1968spherical}. However, the assumption of translational invariance is necessary to obtain the single constraint equation of the spherical model. On general graphs the analysis is considerably more intricate \cite{khoruzhenko1989large,dantchev2014casimir,knops1973infinite}. An important result is that the universality class of the large-$n$ model on graphs is the same as the spherical one \cite{burioni2000limit} and depends only on the spectral dimension \cite{cassi1999spherical}.

In this paper, we characterize the large-$n$ limit of continuous spins on general, nonrandom graphs and we show that the limiting model can be formulated directly in terms of the Laplacian and Adjacency matrices of the underlying graph. The importance of determining and characterizing the model obtained in the large-$n$ limit for graphs is that with this one can perform analytical and numerical calculations of equilibrium and dynamical quantities in a much easier way than directly dealing with the original $O(n)$ problem. The reason is that the limiting model is Gaussian. Whether the large-$n$ approximation gives good results for finite $n$ depends on the problem at hand, but it is very useful both in giving first qualitative results and as a basis of an $1/n$ expansion.

\section{The Laplacian and Adjacency matrices} 
Let us introduce the two main characters of this paper. For a connected graph $\mathcal{G}$, the Adjacency matrix $A$ is given by:
\begin{equation}
    A_{ij}=\begin{cases}
        1&\text{if sites $i$ and $j$ are connected,}\\
        0&\text{otherwise,}
    \end{cases}
\end{equation}
while the Laplacian matrix $L$ is defined as:
\begin{eqnarray}
    L_{ij}=\delta_{ij}z_i-A_{ij},
\end{eqnarray}
where $z_i=\sum_jA_{ij}$ is the number of nearest neighbors, i.e., the coordination number of site $i$. The spectral properties of both of these matrices play a central role in graph theory \cite{harary1969graph,bondy1976graph,merris1994laplacian} and determine many physical properties of the models defined on their corresponding graphs. The Laplacian is positive semidefinite with a single zero eigenvalue and a corresponding constant eigenvector: the ground state is homogeneous in {\it any} graph. At variance, the ground state of the Adjacency matrix tends to localize around sites with a larger coordination number. On translationally invariant lattices these matrices share the same spectrum up to a constant shift ($z=z_i$, for each site $i$), while for general graphs the difference between their low-energy properties gives rise to distinct physical properties related to the localization of particles or modes around the more connected sites   \cite{burioni2001bose,brunelli2004topology,burioni2005random,millan2021complex}. The question we aim to answer is whether, on a general graph in the large-$n$ limit, we find the Laplacian or the Adjacency matrix. Briefly, the answer is both. 

We  show that the free energy of the $O(n)$ model in the large-$n$ limit is governed by the Laplacian and Adjacency spectra in different temperature limits. The Laplacian is retrieved for temperature $T \to 0$, while the Adjacency matrix for $T \to \infty$. We study the Laplacian-to-Adjacency interplay and we present analytical solutions of the saddle-point equations for several representative graphs.

\section{The Model}
The ferromagnetic classical $O(n)$ model on a graph $\mathcal{G}$, with $N$ sites, is defined by the Hamiltonian:
\begin{equation}
    -\beta H=K\sum_{i,j=1}^NA_{ij}\mathbf{S}_i\cdot \mathbf{S}_j=K\sum_{i,j=1}^N\sum_{a=1}^nA_{ij}S_{i,a}S_{j,a},\label{ONmodel}
\end{equation}
where $K=\beta J>0$ is a dimensionless, ferromagnetic coupling and $\mathbf{S}_i$ are n-component vectors with norm $\mathbf{S}_i\cdot \mathbf{S}_i=n$ (with $k_B=1$ and $\beta=1/T$). This normalization is chosen to correctly define the large-$n$ limit for regular lattices \cite{stanley1968spherical}. 

A key observation is that the model is written purely in terms of the Adjacency matrix. However, one can also write the partition function $Z$ associated to Eq. \eqref{ONmodel} in the following way, using the integral representation of the delta function:
\begin{align}
    Z=&\int\prod_{i=1}^N\prod_{a=1}^n dS_{i,a}\,\frac{1}{(2\pi i)^N}\int_{-i\infty}^{i\infty}\prod_{i=1}^Nd\lambda_i
    \notag\\&\exp\left(K\sum_{i,j=1}^N\sum_{a=1}^nA_{ij}S_{i,a}S_{j,a}+\sum_{i=1}^N \lambda_i(1-S_{i,a}^2)\right).
\end{align}
As the individual components $a$ are independent of each other this is equivalent to dropping the index and writing:
\begin{align}
    Z=&\frac{1}{(2\pi i)^N}\int_{-i\infty}^{i\infty}\prod_{i=1}^Nd\lambda_i\left\{\int\prod_{i=1}^N dS_{i}\exp\left(K\sum_{i,j}A_{ij}S_{i}S_{j}\right.\right.\notag\\
    &\left.\left.+\sum_i \lambda_i(1-S_{i}^2)\right)\right\}^n.\label{partition}
\end{align}
In the seminal paper \cite{stanley1968spherical}, Stanley proved that on regular lattices the free energy corresponding to Eq. \eqref{partition} in the limit $n\rightarrow\infty$ is equal to the free energy of the spherical model. To see this, one makes use of a saddle point approximation on Eq. \eqref{partition}, which turns the integration over $\lambda_i$ into a constraint equation. The Hamiltonian of the so-called 
{\it mean} spherical model is then given by:
\begin{equation}
    -\beta H = \sum_{i,j=1}^NS_i(KA_{ij}-\delta_{ij}\lambda)S_j,
\end{equation}
where $S_i$ is a real valued scalar, and the Lagrange multiplier $\lambda$ is chosen such that the average spherical constraint $\langle\sum_i\ S_i^2\rangle=N$ is fulfilled. The fact that only a single Lagrange multiplier is required when taking the limit arises from the equivalence of all lattice sites in a regular lattice. It is known \cite{burioni2000limit,knops1973infinite}, that when performing the large-$n$ limit on more general graphs, it is \emph{a priori} necessary to include as many Lagrange multipliers as there are sites such that the correct Hamiltonian in this case reads:
\begin{equation}
    -\beta H = \sum_{i,j=1}^NS_i(KA_{ij}-\delta_{ij}\lambda_i)S_j \equiv -\sum_{i,j=1}^NS_iM_{ij}S_j,\label{Hamiltonian}
\end{equation}
The partition function of this model is simply given by:
\begin{equation}
    Z=\sqrt{\frac{\pi^N}{\det M}}.
\end{equation}
The complexity lies in the Lagrange multipliers $\lambda_i$, which have to be determined by enforcing $\langle S_i^2\rangle=1,\; \text{for } i=1,\dots,N$. In the thermodynamic limit one is therefore left with infinitely many constraint equations, so that explicit solutions are rare \cite{khoruzhenko1989large,dantchev2014casimir}. It should be noted that if all $\lambda_i$ are equal, the free energy connected to Eq. \eqref{Hamiltonian} will be determined by the eigenvalues of the Adjacency matrix shifted by a constant. If, on the other hand, $\lambda_i=K z_i$ one obtains the Laplacian matrix. Whenever all sites of the graph have the same coordination number, this difference disappears.

\begin{figure}
    \centering
    \includegraphics[width=0.5\linewidth]{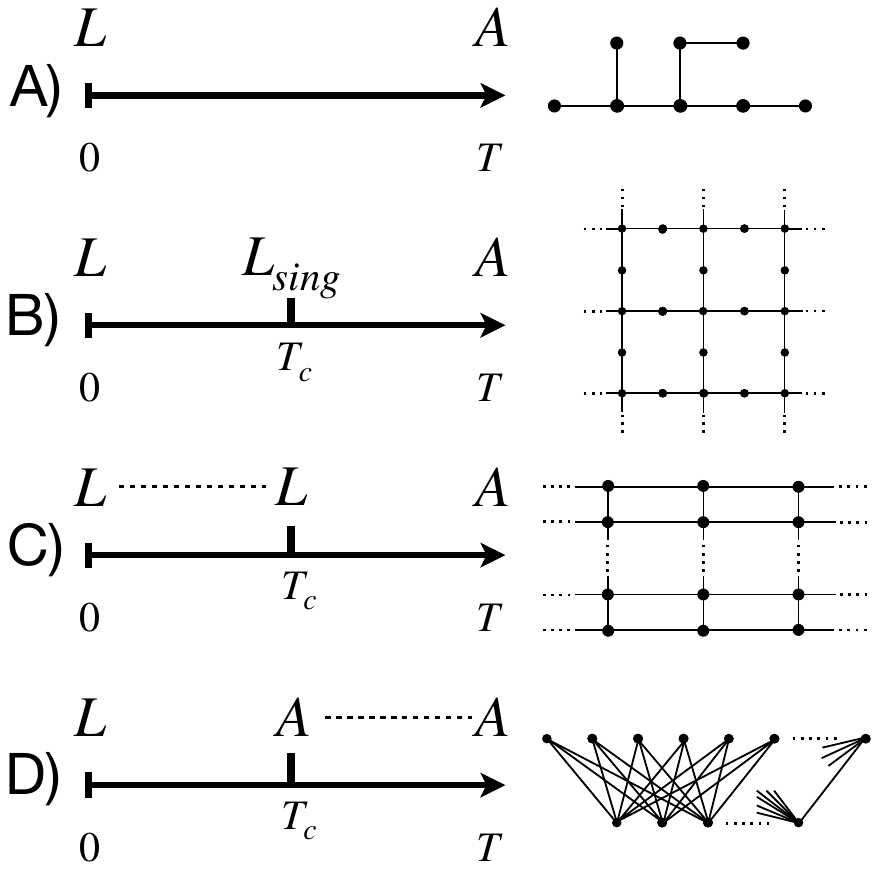}
    \caption{Appearance of the Laplacian (denoted in the figure as "$L$") and Adjacency ("$A$") matrices for the graphs considered in the text. A) For trees there is no critical point such that one only has the high-and low-temperature limit. B) In the decorated lattice the singular part of the free energy is determined by the Laplacian matrix (denoted by "$L_{sing}$"). C) For the strip the Laplacian matrix is retrieved in the whole low temperature regime. D) The complete bipartite graph has diverging coordination numbers and therefore does not have to obey the result in \protect\cite{burioni2000limit}, instead the transition is governed by the Adjacency matrix.}
    \label{figure1}
\end{figure}

\section{The High- and Low-Temperature Limit}
The Lagrange multipliers are determined by a set of saddle point equations for the free energy density $f$:
\begin{eqnarray}
    \frac{\partial}{\partial \lambda_i}f=0,\quad \text{for }i=1,\dots,N ,\label{saddle_point}
\end{eqnarray}
where as usual $\beta f=-\log{Z}/N$ (from now on we will refer to $f$ as simply the free energy). In general, a closed-form solution cannot be obtained since Eq. \eqref{saddle_point} requires, at minimum, knowledge of the determinant of $M$ and its derivatives.

In the low temperature limit ($K\rightarrow\infty$), the matrix $M$ is governed by the ground state of Eq. \eqref{Hamiltonian}, which can be found by solving:
\begin{equation}
    0=2K\sum_{j\in \partial i}S_j-2\lambda_i S_i +h,\qquad0=1-S_i^2,
\end{equation}
where $\partial i$ is the set of sites connected to $S_i$ and $h>0$ a homogeneous external magnetic field that was added to the Hamiltonian \eqref{Hamiltonian} to obtain a unique ground state. The minimum energy solution is then given by:
\begin{eqnarray}
    S_i=1,\quad \lambda_i=Kz_i+\frac{h}{2}.
\end{eqnarray}
In the absence of a magnetic field, the Lagrange multipliers are such that $M$ is exactly proportional to the Laplacian matrix:
\begin{equation}
    \lambda_i=Kz_i,\quad M_{ij}=KL_{ij}.\label{lowT}
\end{equation}
The eigenvalues of the matrix $M$, which determine the free energy, are therefore given by the ones of the Laplacian matrix and the Lagrange multipliers are solely determined by the number of connections per site. 

For high temperatures i.e., $K\rightarrow 0$, one can calculate the partition function perturbatively by simply expanding the determinant of $M$:
\begin{align}
    \det M&=\det(\text{diag}(\lambda_1,\dots,\lambda_N)-K A)\\
    &= \prod_{i=1}^N\lambda_i\left(1+K\Tr [A\cdot\text{diag}(1/\lambda_1,\dots,1/\lambda_N)]+\mathcal{O}(K^2)\right)\\
    &=\prod_{i=1}^N\lambda_i+\mathcal{O}(K^2).
\end{align}
The saddle point equations in terms of the determinant are given by:
\begin{equation}
    \det M=\frac{1}{2}\frac{\partial}{\partial \lambda_i}\det M,
\end{equation}
such that the solution to second order simply reads:
\begin{equation}
    \lambda_i=\frac{1}{2}+\mathrm{O}(K^2),\quad M_{ij}=\frac{1}{2}\delta_{ij}-KA_{ij},\label{highT}
\end{equation}
implying that the eigenvalues of the Adjacency matrix, shifted by a large constant such that all of them are positive, determine the free energy. The Lagrange multipliers become independent of the graph structure. 

The crossover between Laplacian and Adjacency matrix in this model has direct consequences for the ground state (i.e. the eigenvector of $M$ associated to the lowest eigenvalue). The Laplacian forces the ground state to be homogeneous in the low temperature limit, even when sites with high coordination number are present. As temperature is increased one can then observe the effect of localization in the ground state as it approaches the Adjacency matrix. This can be compared to the problem of non-interacting bosons on graphs, where condensation follows a similar mechanism as in the spherical model. The chemical potential fulfills the role of a global Lagrange multiplier leading to an Adjacency matrix spectrum at any temperature. As a consequence, one can observe localized condensation around sites with high coordination number \cite{brunelli2004topology}.

These considerations already capture the essential role of graph topology in both temperature limits. The remaining question concerns the behavior between the two limits and, if it exists, at the critical point. In \cite{burioni2000limit} it was shown that the singular part of the free energy associated with the Hamiltonian in Eq. \eqref{Hamiltonian} coincides with that of a modified spherical model on the same graph, characterized by a single Lagrange multiplier  \cite{cassi1999spherical}:
\begin{equation}
    \tilde{M}_{ij}=-KA_{ij}+\delta_{ij}z_i\lambda,
\end{equation}
such that the Lagrange multiplier $\lambda$ enforces a generalization of the lattice constraint to graphs $\langle\sum_iz_i S_i^2\rangle=N$.
The existence of a critical point is tied to the spectral dimension $d_s$ of the graph \cite{cassi1999spherical}. Graphs with $d_s>2$ exhibit a phase transition in the spherical model and, therefore, also in the large-$n$ limit of the $O(n)$ model.
At the critical temperature $\lambda^{(c)}=K_c$. However, it is not true that the Lagrange multipliers $\lambda_i$ of the model obtained in the large-$n$ limit satisfy $\lambda_i^{(c)}=z_i \lambda_c$ at the critical point. The following sections will illustrate the behavior of the Lagrange multipliers for different representative graphs. Our findings are summarized in Fig. \ref{figure1}, starting from trees that are the subject of the next section.

\section{Trees} 
Trees are graphs without loops which makes their saddle point equations analytically tractable. In a tree (with more than one node), there exists always at least one site $S_k$ with coordination number $z_k=1$, also called a leaf node. 
\begin{align}
    Z=&\int \prod_{i=1}^N dS_i \exp\left(-\sum_{i,j\neq k}^NS_i(\delta_{ij}\lambda_i-KA_{ij})S_j\right.\notag\\
    &\left.+\sum_{i\neq k}^N\lambda_i-\lambda_kS_k^2+2KS_kS_{k'}+\lambda_k\right).
\end{align}

By integrating this site out and shifting the Lagrange multiplier of the neighboring site $\lambda_{k'}$ according to:
\begin{equation}
    \tilde{\lambda}_i=\lambda_{i}-\delta_{ik'}\frac{K^2}{\lambda_k},
\end{equation}
one can isolate all terms depending on $\lambda_k$ in the partition function:
\begin{align}
    Z=&\sqrt{\frac{\pi}{\lambda_k}}e^{\lambda_k+\frac{K^2}{\lambda_k}}\int \prod_{i\neq k} dS_i \notag\\&\exp\left(-\sum_{i,j\neq k}S_i\left(\delta_{ij}\tilde{\lambda}_i-KA_{ij}\right)S_j+\sum_{i\neq k}\tilde{\lambda}_i\right).
\end{align}
This leads to the saddle point equation for $\lambda_k$:
\begin{equation}
    \frac{1}{2\lambda_k}=-\frac{K^2}{\lambda_k^2}+1,
\end{equation}
All leaf nodes of the original tree share therefore the same value for their associated Lagrange multipliers:
\begin{equation}
    \lambda_k^{\text{leaf}}=\frac{1}{4}\left(1+\sqrt{1+16K^2}\right).
\end{equation}

\begin{figure}
    \centering
    \includegraphics[width=0.5\linewidth]{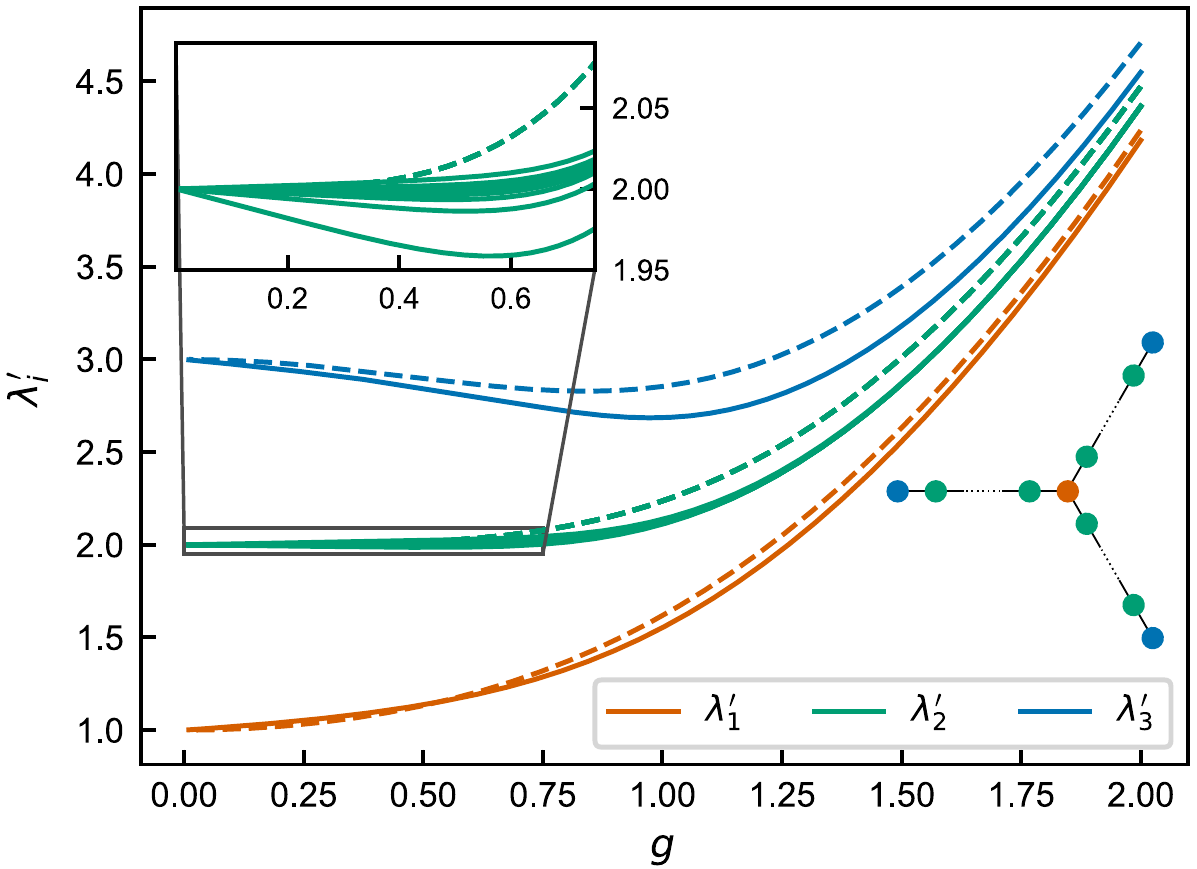}
    \caption{Saddle point values of the rescaled Lagrange multipliers $\lambda_i'=g\lambda_i$ corresponding to sites with $i$ neighbors of the Quantum Rotor model at $T=0$ in the large-$n$ limit on a $Y$ graph (solid lines) as a function of the coupling $g$ (defined as in \protect\cite{vojta1996quantum}). The solution depends on the position inside the leg of the junction. In contrast the classical model only depends on the number of neighbors (dashed lines). Temperature $T$ in the classical model can be identified with $g^2$ in the quantum model.}
    \label{figure2}
\end{figure}

The only effect on the other sites is a renormalization of the neighboring Lagrange multiplier. The remaining graph is still a tree and one can therefore repeat this procedure iteratively for all sites. In the end, the Lagrange multipliers for any tree depend solely on the number of nearest neighbors and noticeably take the simple final form:
\begin{equation}
    \lambda_{i}=\frac{1}{2}+\frac{z_i}{4}\left(-1+\sqrt{1+16K^2}\right).\label{constr_tree}
\end{equation}
The fact that the constraint equations can be solved for any tree is one of the main results of the paper, extending the classical result \cite{stanley1969exact}
of the $O(n)$ model on an open chain. Tree graphs explicitly highlight how both the Laplacian and the Adjacency matrix naturally emerge from the structure of $M_{ij}$:
\begin{equation}
    M_{ij}\sim \begin{cases}
			KL_{ij}, & \text{for $K\rightarrow\infty$},\\
            -KA_{ij}+\frac{1}{2}\delta_{ij}, & \text{for $K\rightarrow 0$}.
		 \end{cases}
\end{equation}
The local nature of the Lagrange multipliers is special to trees and does not generalize to most other graphs. Promoting the $O(n)$ model to the Quantum Rotor model \cite{vojta1996quantum} and taking the large-$n$ limit, this feature is lost as well (see Appendix \ref{app:quantum} for more details). We illustrate this in Fig. \ref{figure2}, where the behavior of the Lagrange multipliers of both models on a $Y$ graph is compared. In the classical model the Lagrange parameters $\lambda_i$ only differ in sites with distinct $z_i$. In the quantum case, however, 
the $\lambda_i$ of sites with the same coordination number (the sites within the chains of the $Y$ structure) are different.
Nevertheless, at high temperature they both tend to converge and reproduce the value $\lambda_i$ of the classical model and their sole dependence on $z_i$. Moreover, also for $T \to 0$ they approach the same value, to reproduce a behavior determined solely by the Laplacian matrix. 

\section{Decorated Lattices} 
\begin{figure}
    \centering
    \includegraphics[width=0.5\linewidth]{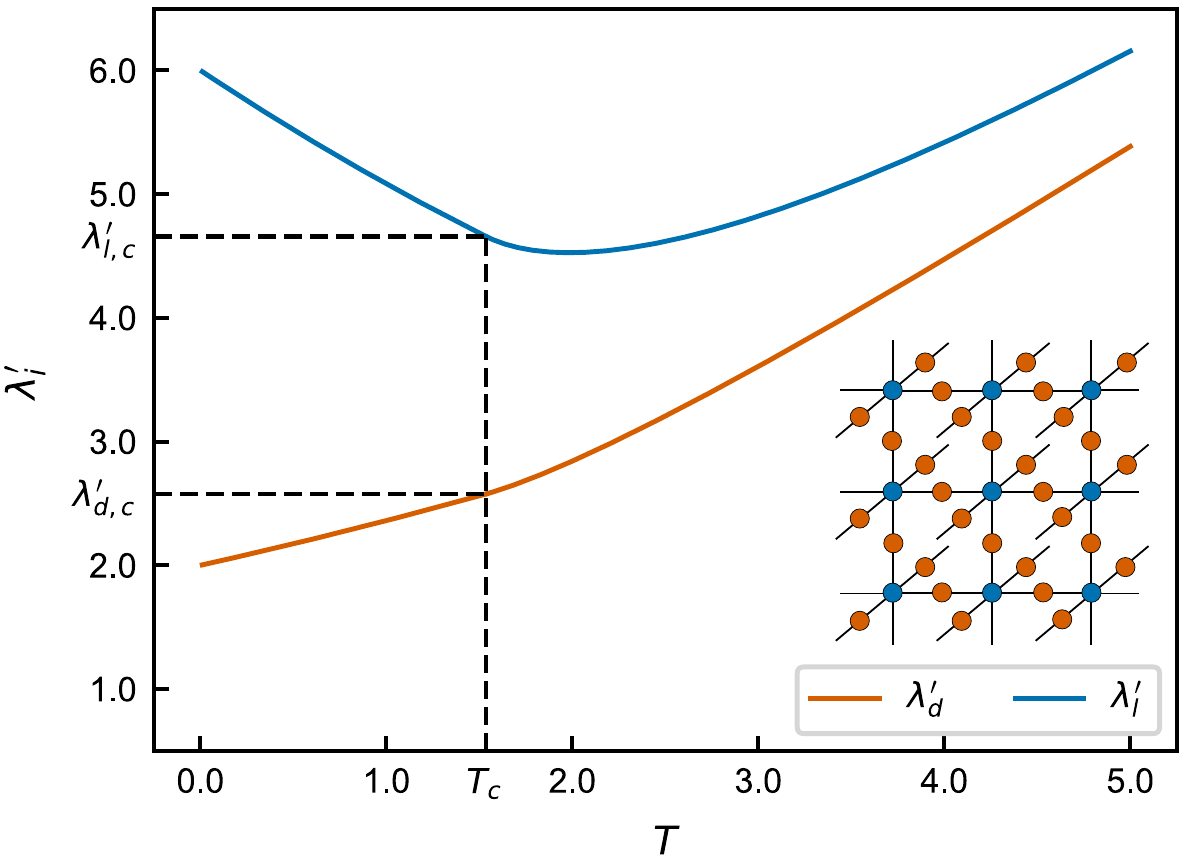}
    \caption{Saddle point values of the rescaled Lagrange multipliers $\lambda_i'=T\lambda_i$ on a decorated lattice in three dimensions. Lagrange multipliers corresponding to decorating sites are denoted by $\lambda_d$, while the ones associated to the sites of the underlying lattice by $\lambda_l$. Above $T_c$, Eqs. \eqref{decor_constraint1} and \eqref{decor_constraint2} must be used, while below $T_c$ one has to minimize \eqref{decorated_below_tc}.}
    \label{figure3}
\end{figure}

A simple way to partially break translational invariance is to consider hypercubic lattices with decorations. This has been studied in \cite{khoruzhenko1989large} focusing on the critical temperature $T_c$ as the number of decorations between sites grows. For simplicity, we consider here just one decoration in $d=3$ dimensions such that every site has either two or 6 neighbors, but the calculations can easily be generalized. There are two types of sites: decorating sites, with Lagrange multiplier $\lambda_d$, and bulk sites of the underlying cubic lattice with Lagrange multiplier $\lambda_l$. Every decorating spin $S_d$ is connected to two spins of the underlying cubic lattice such that it can be integrated out. In the thermodynamic limit the two constraint equations read \ref{app:decorated}:
\begin{align}
    K&=\frac{\lambda_d}{2}\int\frac{d^3k}{(2\pi)^3}\frac{1}{\frac{\lambda_d\lambda_l}{2}-3-\sum_{i=1}^3 \cos k_i}\label{decor_constraint1},\\
    3 K&=\frac{2}{\lambda_d}+\frac{\lambda_l}{2}\int\frac{d^3k}{(2\pi)^3}\frac{1}{\frac{\lambda_d\lambda_l}{2}-3-\sum_{i=1}^3 \cos k_i},\label{decor_constraint2}
\end{align}
where the Lagrange multipliers have been rescaled by $K$ compared to the definition in the previous section, and the integration is over the first Brillouin zone of the underlying cubic lattice, $k_i\in [-\pi,\pi]$. The critical point corresponds to largest value of $K$ for which \eqref{decor_constraint1} and \eqref{decor_constraint2} have a solution. The critical temperature $K_c$ is finite in this case because the integrals do not diverge as the denominator approaches zero. It is easy to see that at $K=K_c$ one needs to have $\lambda^{(c)}_l\lambda_d^{(c)}=12$. One can also convince themselves that the solutions to \eqref{decor_constraint1} and \eqref{decor_constraint2} are different from the Laplacian ones:
$\lambda^{(c)}_l\neq 6$ and $\lambda_d^{(c)}\neq 2$.
In the standard spherical model, the Lagrange multiplier sticks to its critical value below $T_c$. However, if one has more than one parameter, this is no longer the case since only the condition for criticality, in this case $\lambda_l\lambda_d=12$, must be satisfied. The free energy must be minimized, under this constraint. The formerly non-analytic part is absorbed in the constant $A$ and the free energy can be expressed as:
\begin{equation}
    -\beta f=2\ln\lambda_d-3\lambda_dK-\frac{12K}{\lambda_d}+A.\label{decorated_below_tc}
\end{equation}
Minimizing the free energy with respect to $\lambda_d$ gives the correct behavior in the $K\rightarrow\infty$ limit. The resulting dependence of $\lambda_l$ and $\lambda_d$ on $K$ is shown in Fig. \ref{figure3}. The two Lagrange multipliers converge for high temperatures and take on a non-trivial value at $T_c$. They approach their Laplacian values $\lambda_l=6$ and $\lambda_d=2$, while maintaining $\lambda_l\lambda_d=12$, in the limit of zero temperature.

{\it Strip---} 
An example where the Lagrange multipliers yield a non-trivial Laplacian matrix in the whole low-temperature phase was studied in \cite{dantchev2014casimir}. There, a three dimensional strip of length $L$ and periodic boundary conditions in the other two directions such that all sites have six neighbors except the boundary sites which have five, was considered. It was shown that, as long as $4\pi(K-K_c)\gg \frac{\log L}{L}$, where $K_c$ is the bulk critical temperature, is fulfilled, one obtains approximately the Laplacian matrix in the low temperature regime. Therefore, the low temperature expansion is correct even close to $T_c$ as long as $L$ is sufficiently large. This arises from the fact that the fraction of boundary sites approaches zero in the $L\rightarrow\infty$ limit. 

{\it Complete Bipartite Graph---} 
For graphs with finite coordination numbers, the singular part of the model requires the Laplacian matrix \cite{burioni2000limit}. We show that if one relaxes this condition, one can obtain a surprising behavior, namely that the critical behavior is determined by the Adjacency matrix. To illustrate this, we choose a bipartite graph where subset $A(B)$ has $N(2N)$ sites. All sites of subset $A$ are connected to all sites of $B$. In this way, the coordination number of each site grows as $2N(N)$. The Hamiltonian is given by:
\begin{equation}
    -\beta H=\frac{K}{N}\left(\sum_{j=1}^{2N}\sum_{i=1}^NS_{i,A}S_{j,B}\right)-\lambda_{A}\sum_{i=1}^NS_{i,A}^2-\lambda_{B}\sum_{i=1}^{2N}S_{i,B}^2,
\end{equation}
where the coupling has been rescaled by $N$ to obtain extensive quantities. The spherical constraints can be solved analytically because of the simple structure of the graph and one arrives at \ref{app:complete}: $\lambda_A=\lambda_B=1/2$ for, $K\leq K_c=1/2$,
implying that the Adjacency matrix determines the behavior in the whole high-temperature phase. Below $T_c$ on the other hand, the solutions approach the correct ratio $\lambda_A/\lambda_B=2$ for $K\to\infty$:
\begin{align}\begin{aligned}
    \lambda_A=&\frac{1}{4}\left(-1+\sqrt{1+64K^2}\right), &&\text{for}\, K\geq K_c,\\
    \lambda_B=&\frac{1}{8}\left(1+\sqrt{1+64K^2}\right), &&\text{for}\, K\geq K_c.
\end{aligned}
\end{align}
Therefore, one still obtains the general structure of Laplacian and Adjacency matrix for this graph with diverging coordination numbers.

\section{Conclusions} 
We have shown that the behavior of the $O(n)$ model in the large-$n$ limit on general, non-random graphs can be understood as an interplay between the Adjacency and Laplacian matrices. At high temperatures, the spectrum of the Adjacency matrix determines the free energy, while at low temperatures the Laplacian spectrum governs it. This behavior is enforced by a set of saddle point equations that simplify in the two temperature limits, where the Lagrange multipliers either approach the local coordination numbers $z_i$ or become equal across sites. This contrasts with the usual definition of the spherical model on graphs, which assumes a single Lagrange multiplier multiplying the matrix $Z_{ij}=z_i\delta_{ij}$. The crossover between Laplacian and Adjacency matrix emerges to be a general property, however, the way it is realized depends on the specific properties of the graph. We illustrated this dependence by analyzing several representative classes of graphs, both with and without a critical point.

We illustrated this result on several representative graphs, both with and without phase transitions. For trees, the Lagrange multipliers were determined analytically, while for decorated lattices we showed how they remarkably approach their Laplacian values in the low-temperature phase. For a graph with diverging coordination number, we demonstrated that the parameters remain fixed at their Adjacency matrix values even at the critical temperature $T_c$.

In this paper, we have mostly studied nonrandom graphs; however, an extension of our results to random graphs would be very interesting to study. In the case of purely ferromagnetic couplings one may expect the Laplacian-to-Adjacency matrix transition to carry over and that behavior similar to scenarios B) and C) of Fig. \ref{figure1} are observed, which would allow to study phenomena related to the localization of eigenvectors in these networks in a direct way \cite{silva2025spectral}. However, for antiferromagnetic or competing interactions a full investigation should be performed. Our work could be also used as a starting point to extend known results concerning the spherical spin glass model and infinite component spin glass models \cite{kosterlitz1976spherical,almeida1978infinite,aspelmeier2004generalized,lupo2017approximating}. The appearance of the Laplacian is tied to a ferromagnetic ground state, while for spin glasses a more complex low-temperature behavior has to be expected for the Lagrange multipliers.

The large-$n$ analysis developed here provides a natural foundation for a $1/n$ expansion on graphs, enabling both the study of critical behavior and the calculation of non-universal quantities. Extending our results to the antiferromagnetic case, especially on graphs that allow frustration, will be particularly interesting. Finally, the same framework can be extended to the quantum regime. In the large-$n$ limit of the $O(n)$ quantum rotor model on graphs, we expect the resulting saddle point equations and Lagrange multipliers again to reveal regimes dominated by the Laplacian or Adjacency spectra.

\acknowledgments
The authors acknowledge funding from the European Union’s Horizon Europe Research and Innovation Programme under the Marie Sk\l{}odowska-Curie Doctoral Network MAWI (Matter-Wave Interferometers) 
under the grant agreement No. 101073088.

\emph{Data availability statement: } Data are available on request to the authors.

\appendix
\section{Large \texorpdfstring{$n$}{n} Limit for the \texorpdfstring{$O(n)$}{O(n)} Quantum Rotor Model on a \texorpdfstring{$Y$}{Y} graph}
\label{app:quantum}
In the main text, we compare the classical $O(n)$ model and the $O(n)$ quantum rotor model in the large $n$ limit on a $Y$ graph. We write the Hamiltonian of the quantum rotor model in the large $n$ limit as \cite{vojta1996quantum}:
\begin{equation}
    H=-\sum_{\langle ij\rangle}\hat{S}_i\hat{S}_j+\frac{g}{2}\sum_i\hat{P}^2_i+\sum_i\lambda_i(\hat{S}_i^2-1).\label{quantumhamiltonian}
\end{equation}
The operators follow canonical commutation relations
\begin{equation}
    [\hat{S}_i,\hat{S}_j]=[\hat{P}_i,\hat{P}_j]=0,\quad [\hat{S}_i,\hat{P}_j]=i\delta_{ij}.
\end{equation}
The free energy associated with Eq. \eqref{quantumhamiltonian} can be straightforwardly calculated:
\begin{equation}
    F=-T\sum_k \ln \sinh\left( \sqrt{\frac{g \Lambda_k}{2}} \frac{1}{T}\right)+\frac{1}{2}\sum_i(\lambda_i-g),
\end{equation}
where $\Lambda_k$ are the eigenvalues of $M$:
\begin{equation}
    M_{ij}=-A_{ij}+\delta_{ij}\lambda_i.
\end{equation}
We restrict ourselves to the zero temperature regime such that the free energy reads:
\begin{equation}
    f=-\sqrt{\frac{g}{2}}\sum_k\sqrt{\Lambda_k}+\frac{1}{2}\sum_i(\lambda_i-g).
\end{equation}
The saddle point equations are obtained through differentiation with respect ot $\lambda_i$:
\begin{equation}
    1=\sqrt{\frac{g}{2}}\sum_k\frac{{\Psi^{2}_{k,i}}}{\sqrt{\Lambda_k}},
\end{equation}
where ${\Psi_{k,i}}$ is the $i$-th component of the $k$-th eigenvector of $M$. The main difference to the classical case is that in general one cannot write the saddle point equations in terms of the determinant, but instead is forced to consider the whole spectrum.

\section{Decorated lattice}
\label{app:decorated}
In the main text, we consider a three dimensional cubic lattice with a single decoration between each site. This is a special case of the decorated lattices considered in \cite{khoruzhenko1989large}. The Hamiltonian reads:
\begin{equation}
    -\beta H = K\sum_{\langle S_l,S_d\rangle}S_lS_d+\sum_{i=1}^N \lambda_l(1-S_{l,i}^2)+\sum_{j=1}^{3N}\lambda_d(1-S_{d,j}^2).
\end{equation}
The first sum runs over all bonds and can be written in this way because there are only connections between decorations $S_d$ and cubic lattice sites $S_l$. We consider the problem of a single decoration because it simplifies the analysis while retaining the essential features we are interested in. Integrating each decorating spin yields an effective cubic lattice with modified interactions and additional diagonal terms. The individual integrals take the form:
\begin{equation}
    \int_{-\infty}^\infty dS_d\exp\left(K(S_{l,1}+S_{l,2})S_d-\lambda_dS_d^2\right)=\sqrt{\frac{\pi}{\lambda_d}}\exp\left[\frac{K^2}{4\lambda_d}(S_{l,1}^2+2S_{l,1}S_{l,2}+S_{l,2}^2)\right].
\end{equation}
After performing all integrals over decorating sites, the partition function becomes
\begin{equation}
    Z=\left(\frac{\pi}{\lambda_d}\right)^{\frac{3N}{2}}e^{N\lambda+3N\lambda_{d}}\int\prod_{i=1}^NdS_{l,i}\exp\left[\sum_{\langle i,j\rangle}\frac{K^2}{2\lambda_d}S_{l,i}S_{l,j}-\sum_{i=1}^N\left(\lambda_l-\frac{3K^2}{2\lambda_d}\right)S_{l,i}^2\right],
\end{equation}
where $\langle i,j\rangle$ now denotes to the usual nearest neighbor sum on a cubic lattice. The remaining integrals can easily be evaluated by comparison with the known result for the three dimensional cubic lattice. After rescaling $\lambda_l$ and $\lambda_d$ by $K$, the free energy in the thermodynamics limit reads
\begin{equation}
    -\beta f=2\ln\lambda_d-3\lambda_dK-\lambda K+\int\frac{d^dk}{(2\pi)^d}\ln\left(\frac{\lambda_d\lambda_l}{2}-3-\sum_{i=1}^3 \cos k_i \right)+const.
\end{equation}
By taking derivatives with respect to $\lambda_l$ and $\lambda_d$ one obtains the saddle point equations (27) and (28). At the critical point $\lambda^{(c)}_d\lambda^{(c)}_l=12$ and the equations read:
\begin{equation}
    K_c=\frac{\lambda^{(c)}_d}{2}K_{c,0},\qquad
    3 K_c=\frac{2}{\lambda^{(c)}_d}+\frac{\lambda^{(c)}_l}{2}K_{c,0},\qquad
    \lambda^{(c)}_d\lambda^{(c)}_l=12,
\end{equation}
where $K_{c,0}$ is the critical coupling of the three dimensional model without decoration and is given by a finite Watson integral:
\begin{equation}
    K_{c,0}=\int\frac{d^3k}{(2\pi)^3}\frac{1}{3+\sum_{i=1}^3 \cos k_i}.
\end{equation}
Solving the above system for $K_c$, $\lambda_l$ and $\lambda_d$ gives:
\begin{equation} 
K_c = \frac{\sqrt{K_{c,0}} \sqrt{1 + 3 K_{c,0}}}{\sqrt{3}},\qquad
\lambda_l^{(c)} = \frac{6\sqrt{3K_{c,0}}}{\sqrt{1 + 3 K_{c,0}}},\qquad
\lambda_d^{(c)} = \frac{2 \sqrt{1 + 3K_{c,0}}}{\sqrt{3} \sqrt{K_{c,0}}}.
\end{equation}  
Below $T_c$, we impose the condition $\lambda_l\lambda_d=12$ to express the free energy in terms of a single Lagrange multiplier. Up to a constant $A$, the free energy reads:
\begin{equation}
    -\beta f=2\ln\lambda_d-3\lambda_dK-\frac{12K}{\lambda_d}+A.
\end{equation}
The saddle point equation below $T_c$ is given by:
\begin{equation}
    0=\frac{2}{\lambda_d}-3K+\frac{12K}{\lambda_d^2},
\end{equation}
with the solution:
\begin{eqnarray}
    \begin{aligned}
    \lambda_l&=\frac{-1 + \sqrt{1 + (6K)^2}}{K},&&\quad \text{for }K>K_c,\\
    \lambda_d&=\frac{1+\sqrt{1+(6K)^2}}{3K},&&\quad \text{for }K>K_c.
    \end{aligned}
\end{eqnarray}
The above expressions approach the expected Laplacian values $\lambda_l=6$ and $\lambda_d=2$ for $K\rightarrow\infty$.

\section{Complete bipartite graph}
\label{app:complete}
We divide the nodes into two subsets $A$ and $B$ such that there are $N$ nodes in $A$ and $2N$ nodes in $B$. Afterwards all sites belonging to different subsets are connected. The Hamiltonian is given by:
\begin{equation}
    -\beta H=\frac{K}{N}\left(\sum_{j=1}^{2N}\sum_{i=1}^NS_{i,A}S_{j,B}\right)+\lambda_{A}\sum_{i=1}^N(1-S_{i,A}^2)+\lambda_{B}\sum_{i=1}^{2N}(1-S_{i,B}^2).
\end{equation}
The couplings are rescaled by $N$ to obtain extensive thermodynamic quantities. The matrix $M$ has the following structure:
\begin{eqnarray}
    M &&= \begin{pmatrix} 
    D_A && -\frac{K}{N} J_{N\times 2N} \\
    -\frac{K}{N} J_{2N\times N} && D_B
    \end{pmatrix},\\
    D_A&&=\mathrm{diag}(\lambda_A,...,\lambda_A),\\
    D_B&&=\mathrm{diag}(\lambda_B,...,\lambda_B),
\end{eqnarray}
where $J_{N\times M}$ is the $N\times M$ matrix with all elements equal to one. The determinant of $M$ can be calculated by using first a identity for block matrices:
\begin{equation}
    \det\begin{pmatrix} 
    A & B \\
    C & D
    \end{pmatrix}=\det{(A)}\det{\left(D-CA^{-1}B\right)},
\end{equation}
and then the determinant identity for $n\times n$ matrices of the form:
\begin{equation}
    \det\begin{pmatrix} 
    a & b &\dots  &b \\
    b & a & \ddots&\vdots\\
    \vdots  &   \ddots     & \ddots& b\\
    b&\dots &b &a
    \end{pmatrix}=(a+(n-1)b)(a-b)^{n-1}
\end{equation}
In the end, the determinant reads:
\begin{equation}
    \det{M}=\lambda_A^{N-1}\lambda_B^{2N-1}\left(\lambda_A\lambda_B-2K^2\right).
\end{equation}
This gives rise to the following saddle point equations:
\begin{eqnarray}
    N&=\frac{N}{2\lambda_{A}}+\frac{\lambda_B}{\lambda_A\lambda_B-2K^2},\\
    2N&=\frac{N}{\lambda_{B}}+\frac{\lambda_A}{\lambda_A\lambda_B-2K^2}.
\end{eqnarray}
In the $N\rightarrow\infty$ limit, these equations are solved by:
\begin{equation}
    \lambda_A=\lambda_B=\frac{1}{2},\quad \mathrm{for}\, K< K_c=\frac{1}{2\sqrt{2}}.
\end{equation}
As $K\rightarrow K_c$ one approaches a zero of the partition function. The correct way to proceed in the low temperature regime, is to fix $\lambda_A\lambda_B-2K^2$ and minimize the regular part. This leads to the following equations:
\begin{eqnarray}
    0&=\lambda_A\lambda_B-2K^2,\\
    0&=\frac1{2\lambda_A}+1-\frac{4K^2}{\lambda_A}.
\end{eqnarray}
The solutions are:
\begin{eqnarray}
    \lambda_A&=\frac{1}{4}\left(-1+\sqrt{1+64K^2}\right)\\
    \lambda_B&=\frac{1}{8}\left(1+\sqrt{1+64K^2}\right).
\end{eqnarray}
At the critical coupling $K_c$ they are equal to $1/2$ and for $K\rightarrow\infty$ their ratio approaches $2$ as expected.

\bibliography{bibliography_epl}

\end{document}